\documentclass[11pt,fleqn]{article}
\usepackage{amssymb,latexsym,amsmath,amsfonts,graphicx, ebezier}

    \advance\textheight by \topskip

    \numberwithin{equation}{section}

    \def\Re{{\rm Re \,}}
    \def\Im{{\rm Im \,}}
    \def\Ai{{\rm Ai \,}}

    \def\sech{{\rm sech}}

    \def\bigO{{\cal O}}

    \def\P2n{{\rm P}_{{\rm II}}^{(n)}}

    \newtheorem{theorem}{Theorem}[section]
    
    \newtheorem{corollary}[theorem]{Corollary}
    \newtheorem{proposition}[theorem]{Proposition}
    
    \newtheorem{Definition}[theorem]{Definition}
    
    \newtheorem{Remark}[theorem]{Remark}
    \newenvironment{remark}{\begin{Remark}\rm}{\end{Remark}}
    \newtheorem{Example}[theorem]{Example}
    
    \newtheorem{Assumptions}[theorem]{Assumptions}

    \newcommand{\e}{\epsilon}

    \newenvironment{proof}%
    {\rm \trivlist \item[\hskip \labelsep{\bf Proof. }]}%
    {\hspace*{\fill}$\Box$\endtrivlist}
    {\rm \trivlist \item[\hskip \labelsep{\bf Proof}]}%
    {\hspace*{\fill}$\Box$\endtrivlist}

    \newcommand{\sgn}{{\operatorname{sgn}}}

\begin{document}
\title{Asymptotics for a special solution to the second member of the Painlev\'e I hierarchy}

\author{T. Claeys}
\maketitle

\begin{abstract}
We study the asymptotic behavior of a special smooth solution
$y(x,t)$ to the second member of the Painlev\'e I hierarchy. This
solution arises in random matrix theory and in the study of
Hamiltonian perturbations of hyperbolic equations. The asymptotic
behavior of $y(x,t)$ if $x\to \pm\infty$ (for fixed $t$) is known
and relatively simple, but it turns out to be more subtle when $x$
and $t$ tend to infinity simultaneously. We distinguish a region of
algebraic asymptotic behavior and a region of elliptic asymptotic
behavior, and we obtain rigorous asymptotics in both regions. We also discuss two critical transitional asymptotic
regimes.
\end{abstract}
\section{Introduction}
We study asymptotics for a solution to the second member in
the Painlev\'e I hierarchy, also called the ${\rm P}_{\rm I}^2$
equation,
\begin{equation}\label{PI2}
    x=ty-\left(\frac{1}{6}y^3+\frac{1}{24}(y_x^2+2yy_{xx})
        +\frac{1}{240}y_{xxxx}\right).
\end{equation}
This is an ODE in the $x$-variable, but the equation depends also on
the parameter $t$.
Given $t$, general solutions to (\ref{PI2}) have an infinite number
of poles in the complex $x$-plane, but in \cite{CV1}, a
special real solution to (\ref{PI2}) was constructed which has no poles for real values of
$x$ and $t$, and which has the asymptotics
\begin{equation}\label{algebr as}
        y(x,t)=\frac{1}{2}z_0 |x|^{1/3}+\bigO(|x|^{-2}),
            \qquad\mbox{as $x\to\pm\infty$,}
    \end{equation}
    for fixed $t\in\mathbb R$,
    where $z_0=z_0(x,t)$ is the real solution of
    \begin{equation}
        z_0^3=-48\sgn(x)+24z_0|x|^{-2/3}t.
    \end{equation}
It is
remarkable but well-known that $y(x,t)$ is also a solution to the KdV equation
\begin{equation}\label{KdV}
y_t+yy_x+\frac{1}{12}y_{xxx}=0.
\end{equation}

In the case $t=0$, the existence and uniqueness of a real solution
$y_0(x)$ with asymptotics (\ref{algebr as}) were proved by Moore in
\cite{Moore}. Thus any real pole-free solution $y(x,t)$ to
(\ref{PI2}) with asymptotics given by (\ref{algebr as}) solves the
Cauchy problem for the KdV equation with initial data
$y(x,0)=y_0(x)$. Since the Cauchy problem for KdV is uniquely
solvable with initial data $y_0$ (see \cite{Menikoff}), $y(x,t)$ can be characterized as the
unique solution to the Cauchy problem for KdV with initial data
$y_0$. The Whitham equations corresponding to this KdV solution were
already studied by Gurevich and Pitaevskii in \cite{GP} and by
Potemin in \cite{Potemin}.

\medskip

The equation (\ref{PI2}) appeared for $t=0$ in a work by Br\'ezin, Marinari, and Parisi \cite{BMP}, where the authors gave
physical arguments supporting the existence of a real pole-free
solution to (\ref{PI2}) (for $t=0$) with asymptotics given by
\begin{equation}\label{asymptotics y}
    y(x)\sim \mp|6x|^{1/3},
        \qquad\mbox{as $x\to\pm\infty$,}
\end{equation} see also \cite{BB}.
The interest in this solution was renewed after Dubrovin
\cite{Dubrovin} conjectured the existence and uniqueness of a real
solution to (\ref{PI2}) which is pole-free for real $x, t\in\mathbb
R$. The uniqueness part of this conjecture remains open until now.
Another part of Dubrovins conjecture suggests that $y(x,t)$
describes the universal asymptotics for Hamiltonian perturbations of
hyperbolic equations near the point of gradient catastrophe for the
unperturbed equation. This family of equations includes among others
the KdV equation and KdV hierarchy, the de-focusing nonlinear
Schr\"odinger equation, and the Camassa-Holm equation. In the
particular case of the KdV equation, the conjecture was proved in
\cite{CG}. Recurrence coefficients for certain critical
orthogonal polynomials (related to unitary random matrix ensembles)
have a similar type of asymptotics involving $y(x,t)$ \cite{BB,
CV2}.

\medskip

In the two above-mentioned applications the ${\rm P}_{\rm I}^2$
solution $y(x,t)$ describes a singular transition (when letting $x$
and $t$ vary) between a region of simple algebraic asymptotics and a
region of more complicated oscillatory asymptotics involving the
Jacobi elliptic $\theta$-function. One may thus expect that $y(x,t)$
itself also exhibits two different types of asymptotics. However in
the asymptotics (\ref{algebr as}), there is no trace of elliptic or
oscillatory asymptotic behavior. The question arises whether
elliptic asymptotics can be observed when letting $x$ and $t$ tend
to infinity simultaneously. In this paper we will prove that indeed,
depending on the precise scaling of $x$ and $t$, $y(x,t)$ admits
either an algebraic or an elliptic asymptotic expansion.

\medskip

\begin{figure}[t]\label{figcusp}
\begin{center}
\includegraphics[scale=0.25]{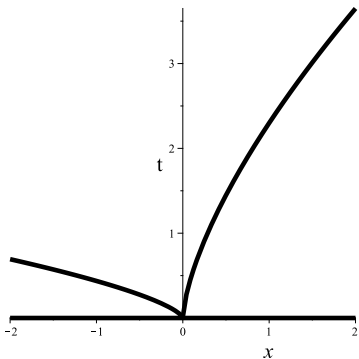}
\end{center}
\caption{$y(x,t)$ has elliptic asymptotics if $(x,t)$ goes to
infinity between the two curves and algebraic asymptotics if $(x,t)$
tends to infinity outside this region.}
\end{figure}

The type of asymptotics we obtain depends on the value of
$s=x|t|^{-3/2}$ as indicated in Figure \ref{figcusp}. If we let
$x\to\pm\infty$, $t \to-\infty$ or if we let $x\to\pm\infty$, $t\to
+\infty$ in such a way that $s=xt^{-3/2}$ remains bounded away from
the interval $[-2\sqrt{3},\frac{2\sqrt{15}}{27}]$, an algebraic
asymptotic expansion similar to (\ref{algebr as}) holds. The leading
term of this expansion will be determined by the equation
$x=ty-\frac{1}{6}y^3$, which is equation (\ref{PI2}) if we ignore
$x$-derivatives. If $x\to\pm\infty$, $t\to +\infty$ in such a way
that $s\in (-2\sqrt{3},\frac{2\sqrt{15}}{27})$ (and remaining
bounded away from the endpoints), the asymptotic expansion for $y$
involves a system of modulation equations and elliptic
$\theta$-functions. This type of behavior was already suggested in
\cite{GP, Potemin}, see also the recent paper \cite{GST}. For $s$
near $-2\sqrt{3}$ and near $\frac{2\sqrt{15}}{27}$, two different
transitions in the asymptotics for $y$ take place.

\begin{remark}
Asymptotics for $y(x,t)$ where both $x$ and $t$ tend to infinity in such a way that $s$ is bounded can
be seen as small dispersion asymptotics for $y$ as a solution to the
KdV equation.  Indeed for $\e>0$, by (\ref{KdV}),
$u(x,t):=\e^{2/7}y(\e^{-6/7}x, \e^{-4/7}t)$ solves the KdV equation
normalized as follows,
\begin{equation}\label{KdV2}
u_t+uu_x+\frac{\e^{2}}{12}u_{xxx}=0.
\end{equation}
\end{remark}

\begin{theorem}{\bf (Algebraic region)}\label{theorem: alg} Let $y$ be the real pole-free solution to the
${\rm P}_{\rm I}^2$ equation defined before. Suppose that either $t<0$, $s\in\mathbb R$ or
$t>0$, $s\in\mathbb R\setminus [-2\sqrt{3},\frac{2\sqrt{15}}{27}]$.
Let $z_0(s)$ be the real zero (which is unique under the above conditions on $s$ and $t$) of the equation
\begin{equation}\label{equation z0}
z_0^3=24\sgn(t)z_0-48s.
\end{equation}
In the limit where $t\to -\infty$, we have
\begin{equation}\label{alg as y}
y(s|t|^{3/2},t)=\frac{z_0(s)}{2}|t|^{1/2}+\bigO(|t|^{-1}),
\end{equation}
uniformly for $s\in\mathbb R$. If $t\to +\infty$ and $s\in\mathbb
R\setminus [-2\sqrt{3},\frac{2\sqrt{15}}{27}]$, the expansion
(\ref{alg as y}) holds as well, uniformly for $s$ bounded away from
the interval $[-2\sqrt{3},\frac{2\sqrt{15}}{27}]$.
\end{theorem}
\begin{remark}
If we only take into account the terms without $x$-derivatives in
(\ref{PI2}), we obtain $x=ty-\frac{1}{6}y^3$. Changing variables
$y\mapsto \frac{z_0}{2}|t|^{1/2}$, this leads to (\ref{equation
z0}). This change of variables will be convenient later on.
\end{remark}
For $s\in (-2\sqrt{3},\frac{2\sqrt{15}}{27})$, the asymptotics for
$y(x,t)$ involve the third Jacobi elliptic $\theta$-function, elliptic integrals, and a system of
modulation equations given by
\begin{align}
&\label{mod1intro}(\beta_3+\beta_2+\beta_1)^2+2(\beta_3^2+\beta_2^2+\beta_1^2)=120,\\
&\label{mod2intro}(\beta_3+\beta_2+\beta_1)^3-4(\beta_3^3+\beta_2^3+\beta_1^3)=360s,\\
&\label{mod3intro}
\int_{\beta_3}^{\beta_2}\sqrt{\xi-\beta_3}(\xi-\alpha)\sqrt{\beta_2-\xi}\sqrt{\beta_1-\xi}d\xi=0,\qquad\qquad
\alpha=-\frac{1}{2}(\beta_3+\beta_2+\beta_1),
\end{align}
where we are interested in solutions
$\beta_3<\alpha<\beta_2<\beta_1$. For the interested reader, we
mention that the variables $\tilde\beta_j=t^{\frac{1}{2}}\beta_j$,
$\tilde\alpha=t^{\frac{1}{2}}\alpha$ solve the elliptic Whitham
equations appearing in the asymptotic theory for the KdV equation
\cite{GT, FRT, V2, W} in the particular case of initial data
\[
\tilde\beta_1(x,0)=\tilde\beta_2(x,0)=\tilde\beta_3(x,0)=-(48x)^{\frac{1}{3}},
\] see \cite{Potemin,
GT}. Those Whitham equations are given by
\begin{equation*}
\label{Whitham} \dfrac{\partial}{\partial
t}\tilde\beta_i+v_i\dfrac{\partial}{\partial x}\tilde\beta_i=0,\quad
v_{i}=\dfrac{2}{3}\frac{\prod_{k\neq
     i}^{}(\tilde\beta_{i}-\tilde\beta_{k})}{\tilde\beta_{i}+a}+\dfrac{1}{3}(\tilde\beta_1+\tilde\beta_{2}+
     \tilde\beta_{3}),\;\;\; i=1,2,3,
\end{equation*}
with
\begin{equation*}
\label{alpha}
a(\tilde\sigma)=-\tilde\beta_{1}+(\tilde\beta_{1}-\tilde\beta_{3})\frac{E(\tilde\sigma)}{K(\tilde\sigma)},\;\;
\;\;
\tilde\sigma^{2}=\frac{\tilde\beta_{2}-\tilde\beta_{3}}{\tilde\beta_{1}-\tilde\beta_{3}},
\end{equation*}
and $K(\tilde\sigma)$, $E(\tilde\sigma)$ are the complete elliptic
integrals of the first and second kind. Solvability of the system
(\ref{mod1intro})-(\ref{mod3intro}) was obtained by Potemin
\cite{Potemin} for $s\in (-2\sqrt{3},\frac{2\sqrt{15}}{27})$, but
can also be decuced from the solvability of the Whitham equations
\cite{FRT, GT}. The asymptotics for $y(x,t)$ in the elliptic region are
expressed in terms of the solution to this system. We obtain
asymptotics of the same type as in \cite{DVZ}, although
only rapidly decaying initial data for KdV were considered there.

\begin{theorem}{\bf (Elliptic region)}\label{theorem: ell}
Let $s\in (-2\sqrt{3},\frac{2\sqrt{15}}{27})$ and let
$\beta_3<\alpha<\beta_2<\beta_1$ solve the system of modulation
equations (\ref{mod1intro})-(\ref{mod3intro}). Then $y(x,t)$ has the following
expansion if we let $t\to +\infty$,
\begin{equation}\label{expansion elliptic intro}
y(st^{3/2},t)=t^{1/2}\frac{\beta_3+\beta_2-\beta_1}{2}+t^{1/2}(\beta_1-\beta_3)\frac{E(\sigma)}{K(\sigma)}+\frac{t^{1/2}}{C^{2}}(\log\theta)''
\left(\frac{t^{7/4}}{2\pi}\Omega\right)+\bigO(t^{-1/2}).
\end{equation}
Here $\theta$ is the third Jacobi $\theta$-function
\begin{equation}
\theta(z;\tau)=\sum_{m=-\infty}^{+\infty}e^{2\pi imz+\pi i\tau m^2},
\end{equation}
and
\begin{align}
&C=\frac{2K(\sigma)}{\sqrt{\beta_1-\beta_3}},\\
&
\Omega=\frac{1}{15}\int_{\beta_2}^{\beta_1}\sqrt{\xi-\beta_3}(\xi-\alpha)\sqrt{\xi-\beta_2}\sqrt{\beta_1-\xi}d\xi,\\
&\tau=\frac{iK'(\sigma)}{K(\sigma)}, \qquad
\sigma=\sqrt{\frac{\beta_2-\beta_3}{\beta_1-\beta_3}},
\end{align}
and $K(\sigma), E(\sigma)$ are the complete elliptic integrals of the first and second
kind. The expansion (\ref{expansion elliptic intro}) holds uniformly
for $s\in (-2\sqrt{3}+\delta,\frac{2\sqrt{15}}{27}-\delta)$ for any
$\delta>0$.
\end{theorem}

If we take asymptotics for $y(x,t)$ where $t\to +\infty$, $x\to
\pm\infty$ in such a way that $s=xt^{-3/2}$ tends to $-2\sqrt{3}$ or
$\frac{2\sqrt{15}}{27}$ sufficiently fast, neither the algebraic
expansion nor the elliptic one is valid. In this case one can obtain
two types of critical asymptotics. Similar transitional asymptotics
have been proved for decaying negative smooth solutions to the KdV equation in \cite{CG2, CG3}. Because
the proofs are very similar as for KdV and for conciseness of the
paper, we will not prove those results here, but we will indicate
below the procedure that can be followed in order to prove them.

\medskip

The first critical regime appears when $s\to -2\sqrt{3}$, which
corresponds to the left edge of the oscillatory zone in the
$(x,t)$-plane, see Figure \ref{figcusp}. The asymptotics involve the
Hastings-McLeod solution \cite{HM} to the second Painlev\'e equation. This is
the unique solution to the Painlev\'e II equation
\begin{equation}
q_{\xi\xi}(\xi)=\xi q(\xi)+2q(\xi)^3,
\end{equation}
with asymptotics given by
\begin{equation}
q(\xi)\sim \Ai(\xi), \quad \mbox{ as $\xi\to +\infty$}, \qquad q(\xi)\sim\sqrt{\frac{-\xi}{2}}, \quad \mbox{ as $\xi\to -\infty$}.
\end{equation}
For $\xi\in\mathbb R$, we have
\begin{equation}\label{expansion P2}
y(-2\sqrt{3}t^{3/2}+c_0t^{1/3}\xi,t)=2\sqrt{3}t^{1/2}-\frac{1}{c_1t^{1/12}}q(\xi)\cos(t^{7/4}\omega)+\bigO(t^{-2/3}),
\qquad\mbox{ as $t\to +\infty$,}
\end{equation}
where $c_0=5^{1/3}$, $c_1=\frac{5^{-1/6}\cdot 3^{-1/4}}{2}$, and
\[\omega=\frac{80}{21}\sqrt{5}\cdot 3^{3/4}-2\cdot 3^{1/4}\cdot 5^{5/6}\xi t^{-1}.\] In other words the leftmost oscillations can be modeled by
a rapidly oscillating cosine and the amplitude develops
proportional to the Hastings-McLeod solution.

\medskip

The second critical regime is the one where $s\to
\frac{2\sqrt{15}}{27}$, i.e.\ the right edge of the oscillatory zone
in the $(x,t)$-plane, see Figure \ref{figcusp}. Here the asymptotic
expansion for $y$ involves a sum of $\sech^2$-terms: we have
\begin{multline}\label{expansion sol}
y\left(\frac{2\sqrt{15}}{27}t^{3/2}-c_2t^{-1/4}\ln t \cdot
\xi,t\right)\\=-\frac{2}{3}\sqrt{15}t^{1/2}+\frac{7}{3}\sqrt{15}t^{1/2}\sum_{k=0}^{\infty}\sech^2(X_k)+\bigO(t^{-5/4}\ln^2
t),
\end{multline}
where $c_2=\frac{\sqrt{7}\cdot 3^{\frac{3}{4}}}{8\cdot 5^{\frac{1}{4}}}$, and
\begin{align}
&\label{def Xk} X_k=-\frac{7}{8}(\frac12+k-\xi)\ln t-\ln(\sqrt{2\pi} h_k)-(k+\frac12)\ln\gamma,\\
&h_k=\dfrac{2^{\frac{k}{2}}
}{\pi^{\frac{1}{4}}\sqrt{k!}},\qquad\qquad
\gamma=4\sqrt{2}\cdot 5^{\frac{3}{8}}\cdot 7^{\frac{5}{4}}\cdot 3^{-\frac{11}{8}}.
\end{align}
Recall that the KdV equation admits soliton solutions of the form
$a\,\sech^2(bx-ct)$. The expansion (\ref{expansion sol}) can thus be
seen as a superposition of solitons. The amplitude of the rightmost
oscillations is of the same order as the leading order term of $y$
outside the oscillatory region, and of the same order as the
amplitude of the oscillations in the elliptic region. Because of the
term with $\ln t$ in (\ref{def Xk}), every soliton is sharply
localized near a positive half integer value of $\xi$.

\subsection{Riemann-Hilbert problem for ${\rm P}_{\rm I}^2$}\label{section:
RHP}

In order to prove our results, we will perform an asymptotic analysis of the Riemann-Hilbert (RH) problem associated to the real pole-free solution to the ${\rm P}_{\rm I}^2$ equation. For a more general description of Riemann-Hilbert problems for Painlev\'e equations we refer to \cite{FIKN}.

 Consider the following RH problem for given
complex parameters $x$ and $t$, on a contour
$\Gamma=\mathbb R^-\cup \bigcup_{j=0}^6e^{\frac{j2\pi i}{7}}\mathbb R^+$ consisting of
eight straight rays orientated from left to right.

\subsubsection*{RH problem for $Y$:}

\begin{itemize}
    \item[(a)] $Y$ is analytic in $\mathbb{C}\setminus\Gamma$.
    \item[(b)] $Y$ satisfies the following jump relations on
    $\Gamma$,
    \begin{align}
        \label{RHP Y: b1}
        Y_+(\zeta)&=Y_-(\zeta)
            \begin{pmatrix}
                1 & s_j \\
                0 & 1
            \end{pmatrix},& \mbox{for $\arg\zeta=\frac{j2\pi}{7}$ with $j$ even,}\\[1ex]
        \label{RHP Y: b2}
        Y_+(\zeta)&=Y_-(\zeta)
            \begin{pmatrix}
                1 & 0 \\
                s_j & 1
            \end{pmatrix},& \mbox{for
            for $\arg\zeta=\frac{j2\pi}{7}$ with $j$ odd,}\\[1ex]
        \label{RHP Y: b3}
        Y_+(\zeta)&=Y_-(\zeta)
            \begin{pmatrix}
                0 & 1 \\
                -1 & 0
            \end{pmatrix},& \mbox{for $\zeta\in\mathbb R^-$.}
    \end{align}
    \item[(c)] $Y$ has an asymptotic expansion of the form
    \begin{equation}\label{RHP Y: c1}
        Y(\zeta)=\left(I+\sum_{k=1}^{\infty}A_k(x,t)\zeta^{-k}\right)\zeta^{-\frac{1}{4}\sigma_3}N
        e^{-\theta(\zeta;x,t)\sigma_3},\qquad \mbox{ as
    $\zeta\to\infty$,}
    \end{equation}
    where
    \begin{equation}\label{definition: N theta}
        N=\frac{1}{\sqrt 2}
        \begin{pmatrix}
             1 & 1 \\
             -1 & 1
        \end{pmatrix}e^{-\frac{1}{4}\pi i\sigma_3},\qquad
        \theta(\zeta;x,t)=\frac{1}{105}\zeta^{7/2}-\frac{1}{3}t\zeta^{3/2}+x\zeta^{1/2},
    \end{equation}
    and $\sigma_3=\begin{pmatrix}1&0\\0&-1\end{pmatrix}$.
\end{itemize}
This is the RH problem for a general solution to the ${\rm P}_{\rm I}^2$ equation (\ref{PI2}). The RH problem can only be solvable if the
Stokes multipliers $s_j$ satisfy the relation
\begin{equation}\label{condition stokes}
\begin{pmatrix}1&0\\-s_4&1\end{pmatrix}
\begin{pmatrix}1&-s_5\\0&1\end{pmatrix}
\begin{pmatrix}1&0\\s_6&1\end{pmatrix}
\begin{pmatrix}1&s_0\\0&1\end{pmatrix}
\begin{pmatrix}1&0\\s_1&1\end{pmatrix}
\begin{pmatrix}1&-s_2\\0&1\end{pmatrix}
\begin{pmatrix}1&0\\-s_3&1\end{pmatrix}
=\begin{pmatrix}0&1\\-1&0\end{pmatrix}.
\end{equation}
For any set of Stokes multipliers $s_0, \ldots, s_6$ (independent of $\zeta, x, t$) satisfying this condition,
it was proved in \cite{CV1} that the function \begin{equation}\label{def y}y(x,t)=2A_{1,11}(x,t)-A_{1,12}^2(x,t),
\end{equation} with $A_1$ given as in (\ref{RHP Y: c1}), is a solution to (\ref{PI2}).
However, $y$ can have poles at
certain isolated values of $(x,t)$. Those points corresponds to the
(isolated) values of $x$ and $t$ at which the RH problem for $Y$ is
not solvable. Since $y$ solves (\ref{PI2}), it also solves the KdV
equation (\ref{KdV}). Even in the more general situation where $s_0,
\ldots, s_6$ depend on $\zeta$ (but not on $x$ and $t$), $y(x,t)$
still solves the KdV equation (locally near $(x,t)$) if the RH problem is solvable at
$(x,t)$.

\begin{figure}[t]
\begin{center}
    \setlength{\unitlength}{1truemm}
    \begin{picture}(100,48.5)(0,2.5)
        \put(50,27.5){\thicklines\circle*{.8}}
        \put(49,30){\small 0}

        \put(50,27.5){\line(-2,1){40}}
        \put(50,27.5){\line(-2,-1){40}}
        \put(5,27.5){\line(1,0){80}}

        \put(76,31){$\Gamma_1$}
        \put(17,48){$\Gamma_2$}
        \put(10,31){$\Gamma_3$}
        \put(17,16){$\Gamma_4$}

        \put(30,37.5){\thicklines\vector(2,-1){.0001}}
        \put(30,17.5){\thicklines\vector(2,1){.0001}}
        \put(25,27.5){\thicklines\vector(1,0){.0001}}
        \put(70,27.5){\thicklines\vector(1,0){.0001}}
    \end{picture}
    \caption{The oriented contour $\Gamma$ consisting of the four straight rays
    $\Gamma_1$, $\Gamma_2$, $\Gamma_3$, and $\Gamma_4$.}
    \label{figure: RHP Psi}
\end{center}
\end{figure}
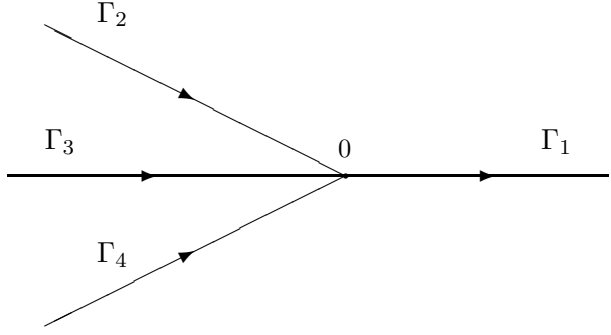

We are interested in one particular solution to the ${\rm P}_{\rm I}^2$ equation. This solution corresponds to the case where
\begin{equation}\label{stokes}
s_1=s_2=s_5=s_6=0, \qquad s_3=s_4=s_0=1,
\end{equation}
which means that there are only jumps for $Y$ on $\Gamma=\cup_{j=1}^4\Gamma_j$, with
\begin{equation}
\Gamma_1=\mathbb R^+, \quad \Gamma_2=e^{\frac{6\pi i}{7}}\mathbb R^+, \quad \Gamma_3=\mathbb R^-, \quad \Gamma_4=e^{-\frac{6\pi i}{7}}\mathbb R^+,
\end{equation}
see Figure \ref{figure: RHP Psi}.
For this choice of Stokes multipliers, it was proved in \cite{CV1} that the RH problem for $Y$ is uniquely solvable for all real values of $x$ and $t$, and that $y$ defined by (\ref{def y}) is the unique real pole-free solution to equation (\ref{PI2}) with asymptotics given by (\ref{algebr as}).
From now on, we refer to $Y$ as the solution to the RH problem with Stokes multipliers given by (\ref{stokes}).
One can verify from the RH conditions that $Y(\zeta)$ has analytic continuations
from each of the four sectors determined by $\Gamma_1, \ldots, \Gamma_4$ to the entire complex plane.
We denote $Y_j$, $j=1, \ldots , 4$ for the analytic continuation from the sector in between
$\Gamma_j$ and $\Gamma_{j\mod 4 +1}$. The jump conditions tell us that
\begin{equation}\label{jumps Y}
Y_1=Y_2\begin{pmatrix}1&0\\1&1\end{pmatrix}, \quad
Y_2=Y_3\begin{pmatrix}0&1\\-1&0\end{pmatrix}, \quad
Y_3=Y_4\begin{pmatrix}1&0\\1&1\end{pmatrix}, \quad
Y_1=Y_4\begin{pmatrix}1&1\\0&1\end{pmatrix}.
\end{equation}

\subsubsection*{Outline}
In the next sections, we will obtain asymptotics for $Y=Y(\zeta;
x,t)$ when $x,t\to\infty$ by applying the Deift/Zhou steepest
descent method \cite{DZ1} on the RH problem. Using (\ref{def y}),
this will also lead to asymptotics for $y(x,t)$. In Section
\ref{section: alg}, we will prove the algebraic expansion for $y$
given in Theorem \ref{theorem: alg}. In Section \ref{section: ell},
we will prove the elliptic asymptotics stated in Theorem
\ref{theorem: ell}. In Section \ref{section: crit}, we will give
a brief overview of the procedure that can be followed to prove the
critical asymptotics (\ref{expansion P2}) and (\ref{expansion sol}). In addition we will make some comments about special solutions to higher members of the Painlev\'e I hierarchy and about unbounded solutions of the KdV equation.

\section{Algebraic region}\label{section: alg}
Here we generalize the asymptotic analysis done in \cite{CV1} for
$t$ fixed, which was inspired by \cite{Kapaev, Kapaev2}. We
assume $t\neq 0$ in what follows.
\subsection{Construction of the $g$-function}
In order to transform the RH problem for $Y$ into a RH problem normalized at infinity and with 'sufficiently simple' jump conditions, we need to construct a so-called $g$-function. This function will be of the following form,
\begin{equation}\label{definition: g}
    g(\zeta)=
        c_1(\zeta-z_0)^{7/2}+c_2(\zeta-z_0)^{5/2}+c_3(\zeta-z_0)^{3/2},
\end{equation}
where we take the branch of the roots which is analytic in $\mathbb C\setminus(-\infty,z_0]$ and positive for $\zeta>z_0$.
Let us write \begin{equation} s=x|t|^{-3/2}.
\end{equation}
If $z_0=z_0(s)$ is the real root (which is unique for $t<0$ and for $t>0$, $s\in\mathbb R\setminus [-2\sqrt{3},\frac{2\sqrt{15}}{27}]$) of the third degree equation
    \begin{equation}\label{definition: z0}
        z_0^3=24\sgn(t)z_0-48s,
    \end{equation}
and if we let
\begin{equation}\label{definition: c1 c2 c3}
    c_1=\frac{1}{105}, \qquad
    c_2=\frac{1}{30}z_0, \qquad
    c_3=\frac{1}{24}z_0^2-\frac{\sgn(t)}{3},
\end{equation}
it is straightforward to check that, with $\theta$ given by
(\ref{definition: N theta}),
\begin{equation}\label{asymptotics g2}
    |t|^{7/4}g(\zeta;s)=\theta(|t|^{1/2}\zeta;x,t)+d_1\zeta^{-1/2}+\bigO(\zeta^{-3/2}),
    \qquad \mbox{as $\zeta\to\infty$,}
\end{equation}
where $d_1$ does not depend on $\zeta$; its explicit value can be calculated but is unimportant.

\begin{proposition}\label{prop g algebr}
Suppose that either $t<0$ or $t> 0$ and $s\in\mathbb R\setminus [-2\sqrt{3},\frac{2\sqrt{15}}{27}]$. Then
\begin{align}
&\label{ineq1}g(\zeta;s)>0,&\mbox{ for
$\zeta>z_0$,}\\
&\label{ineq2}\Im g_+'(\zeta;s)>0,&\mbox{ for $\zeta<z_0$.}
\end{align}
\end{proposition}
\begin{proof}
The function $(\zeta-z_0)^{-3/2}g(\zeta;s)$ is quadratic in $\zeta$
and has no real zeros if $c_2^2-4c_1c_3<0$. Since $c_1>0$, it is
positive for large $\zeta$ and thus for all $\zeta$ if
$c_2^2-4c_1c_3<0$. Now by (\ref{definition: c1 c2 c3}),
\begin{equation}
c_2^2-4c_1c_3=-\frac{1}{315}\left(\frac{3z_0^2}{20}-4\sgn(t)\right),
\end{equation}
which is negative if either $t$ is negative or $t>0$ and
$|z_0|>4\sqrt{\frac{5}{3}}$. If $z_0\in(2\sqrt{2},
4\sqrt{\frac{5}{3}})$, one verifies that both zeros are smaller than
$z_0$ so that (\ref{ineq1}) still holds. In a similar way one shows
that $\Im g_+'(\zeta)$ has no real zeros apart from $z_0$ if
$\frac{25}{4}c_2^2-21c_1c_3<0$, which is true if $t<0$ and if $t>0$,
$|z_0|>4\sqrt{3}$. Moreover in $(-4\sqrt{3}, -2\sqrt{2})$ both zeros
lie to the right of $z_0$ and (\ref{ineq2}) still holds.

We can conclude that (\ref{ineq1}) and (\ref{ineq2}) hold for $t<0$ and for $t>0$ if
$z_0\in\mathbb R\setminus [-4\sqrt{\frac{5}{3}}, 4\sqrt{3}]$. By
(\ref{definition: z0}) this is equivalent to $s\in\mathbb
R\setminus[-2\sqrt{3}, \frac{2\sqrt{15}}{27}]$.
\end{proof}
By (\ref{definition: g}) and the above proposition, a
straightforward complex analysis argument using the Cauchy-Riemann
conditions leads to the following corollary.
\begin{corollary}\label{corollary}
Suppose that either $t<0$ or $t> 0$ and $s\in\mathbb R\setminus
[-2\sqrt{3},\frac{2\sqrt{15}}{27}]$. For $\theta_0>0$ sufficiently
small, we have
\begin{align}
&\Re g(\zeta)<0, &\mbox{ for $\arg(\zeta-z_0)=\pi\pm \theta_0$,}\\
&\Re g(\zeta)>0, &\mbox{ for $\arg(\zeta-z_0)=0$.}
\end{align}
\end{corollary}
These inequalities enable us to transform the RH problem for $Y$ to
a RH problem for which the jumps decay exponentially fast to $I$ as
$t\to \infty$, except on $(-\infty, z_0)$ and in a small
neighborhood of $z_0$.

\subsection{Normalization of the RH problem}
Fix $0<\theta_0<\frac{2\pi}{7}$ such that Corollary \ref{corollary} holds. Let us define $T$ as follows,
\begin{multline}\label{def T}
T(\zeta; x,t)=\begin{pmatrix}1&0\\d_1|t|^{1/4}&1\end{pmatrix}\\
\times\quad
\begin{cases}
Y_1(|t|^{1/2}\zeta; x,t)e^{|t|^{7/4}g(\zeta;x,t)\sigma_3},&\mbox{ if $0<\arg(\zeta-z_0)<\pi-\theta_0$},\\
Y_2(|t|^{1/2}\zeta; x,t)e^{|t|^{7/4}g(\zeta;x,t)\sigma_3},&\mbox{ if $\pi-\theta_0<\arg(\zeta-z_0)<\pi$},\\
Y_3(|t|^{1/2}\zeta; x,t)e^{|t|^{7/4}g(\zeta;x,t)\sigma_3},&\mbox{ if $\pi<\arg(\zeta-z_0)<\pi+\theta_0$},\\
Y_4(|t|^{1/2}\zeta; x,t)e^{|t|^{7/4}g(\zeta;x,t)\sigma_3},&\mbox{ if $\pi+\theta_0<\arg(\zeta-z_0)<2\pi$},
\end{cases}
\end{multline}
with $d_1$ given by (\ref{asymptotics g2}), and with the $Y_j$'s the
analytic extensions of $Y$ as explained in Section \ref{section:
RHP}. Then $T$ satisfies the following RH problem.
\subsubsection{RH problem for $T$}
\begin{itemize}
\item[(a)] $T$ is analytic in $\mathbb{C}\setminus\Sigma$, where $\Sigma=z_0+(\mathbb R\cup e^{\pi\pm\theta_0}\mathbb R^+)$.
    \item[(b)] $T$ satisfies the following jump relations on
    $\Sigma$,
    \begin{align}
        \label{RHP Phi: b1}
        T_+(\zeta)&=T_-(\zeta)
            \begin{pmatrix}
                1 & e^{-2|t|^{7/4}g(\zeta;x,t)} \\
                0 & 1
            \end{pmatrix},& \mbox{for $\arg(\zeta-z_0)=0$,}\\[1ex]
        \label{RHP Phi: b2}
        T_+(\zeta)&=T_-(\zeta)
            \begin{pmatrix}
                1 & 0 \\
                e^{2|t|^{7/4}g(\zeta;x,t)} & 1
            \end{pmatrix},& \mbox{for
            $\arg(\zeta-z_0)=\pi\pm \theta_0$,}\\[1ex]
        \label{RHP Phi: b3}
        T_+(\zeta)&=T_-(\zeta)
            \begin{pmatrix}
                0 & 1 \\
                -1 & 0
            \end{pmatrix},& \mbox{for $\arg(\zeta-z_0)=\pi$.}
    \end{align}
    \item[(c)]  $T$ has the following asymptotic behavior as
    $\zeta\to\infty$,
    \begin{equation}\label{RHP Phi: c}
        T(\zeta)=\left(I+B_1(x,t)\zeta^{-1}+\bigO(\zeta^{-2})\right)
        |t|^{-\frac{1}{8}\sigma_3}\zeta^{-\frac{1}{4}\sigma_3}N,
    \end{equation}
    with
    \begin{equation}\label{def B1}
    B_1=\begin{pmatrix}\frac{d_1^2}{2}+|t|^{-1/2}A_{1,11}-d_1|t|^{-1/4}A_{1,12}&|t|^{-1/2}A_{1,12}-d_1|t|^{-1/4}\\
    *&*\end{pmatrix},
    \end{equation}
    where the values of the *-entries can be calculated but are unimportant for us, and with $A_1$ as in (\ref{RHP Y: c1}).
\end{itemize}
Using (\ref{def y}), we can recover $y(x,t)$ from the identity
\begin{equation}\label{yB}
y(x,t)=2t^{1/2}B_{1,11}-tB_{1,12}^2.
\end{equation}
By Corollary \ref{corollary}, it follows that the jump matrices for $T$ decay to the identity matrix when $|t|\to\infty$ except on $(-\infty, z_0)$ and in a small fixed neighborhood of $z_0$. Ignoring a neighborhood of $z_0$ and ignoring exponential decay, we have jump conditions which can be solved explicitly.

\subsection{Outside parametrix}
We define the outside parametrix by
\begin{equation}\label{def Pinfty alg}
P^{(\infty)}(\zeta)=|t|^{-\frac{\sigma_3}{8}}(\zeta-z_0)^{-\frac{\sigma_3}{4}}N,
\end{equation}
which is analytic in $\mathbb C\setminus (-\infty,z_0]$, and satisfies the jump condition
\[P_+^{(\infty)}(\zeta)=P_-^{(\infty)}(\zeta)\begin{pmatrix}0&1\\-1&0\end{pmatrix}\]
for $\zeta\in(-\infty, z_0)$ if $N$ is given by (\ref{definition: N theta}). Note also that
\begin{equation}
T(\zeta)P^{(\infty)}(\zeta)^{-1}=I+\frac{C_1}{\zeta}+\bigO(\zeta^{-2}),\qquad\mbox{ as $\zeta\to\infty$,}
\end{equation}
with
\begin{equation}\label{def C1}
C_1=B_1-\frac{z_0}{4}\sigma_3.
\end{equation}
The leading order asymptotics for $T$ and $y(x,t)$ will be determined by the outside parametrix.

\subsection{Local parametrix near $z_0$}
In order to obtain asymptotics for $T$ uniformly for
$\zeta\in\mathbb C$, we need to construct a local parametrix in a
small fixed neighborhood $U$ of $z_0$. This local parametrix has
been constructed in \cite[Section 3.4]{CV1} using the Airy function,
and it solves the following RH problem.
\subsubsection*{RH problem for $P$:}
\begin{itemize}
    \item[(a)] $P$ is analytic in $U\setminus\Sigma$.
    \item[(b)] $P_+(\zeta)=P_-(\zeta)v_T(\zeta)$ for $\zeta\in \Sigma\cap U$, where
    $v_T$ is the jump matrix for $T$ given by (\ref{RHP Phi: b1})-(\ref{RHP Phi: b3}).
    \item[(c)] $P(\zeta)P^{(\infty)}(\zeta)^{-1}=I+\bigO(|t|^{-3/2})$, \qquad as $t\to -\infty$ and if $t\to +\infty$ in such a way that $s$ is bounded away from $[-2\sqrt{3},\frac{2\sqrt{15}}{27}]$,
    uniformly for $\zeta\in\partial U$.
\end{itemize}
The local parametrix is needed for the rigor of the RH analysis, but its explicit expression in terms of the Airy function will not be needed; it does not contribute to the leading order asymptotics for $y(x,t)$.

\subsection{Final transformation}
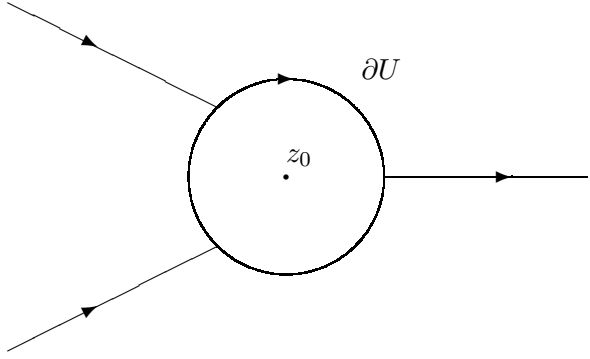
\begin{figure}[t]
\begin{center}
    \setlength{\unitlength}{1truemm}
    \begin{picture}(100,55)(-5,10)
        \cCircle(50,40){13}[f]
        \put(50,40){\thicklines\circle*{.8}}
        \put(50,42){$z_0$}
        \put(60,53){$\partial U$}
        \put(41,49.2){\line(-2,1){28}}
        \put(41,30.8){\line(-2,-1){28}}
        \put(63,40){\line(1,0){27}}
        \put(80,40){\thicklines\vector(1,0){.0001}}
        \put(25,57.2){\thicklines\vector(2,-1){.0001}}
        \put(25,22.8){\thicklines\vector(2,1){.0001}}
        \put(51,53){\thicklines\vector(1,0){.0001}}
    \end{picture}
    \caption{The jump contour $\Sigma_R$.}
    \label{figure: contour R algebr}
\end{center}
\end{figure}

Define
\begin{equation}
R(\zeta;x,t)=\begin{cases}T(\zeta;x,t)P^{(\infty)}(\zeta)^{-1},&\mbox{ for $\zeta\in\mathbb C\setminus U$,}\\
T(\zeta;x,t)P(\zeta)^{-1},&\mbox{ for $\zeta\in U$.}
\end{cases}
\end{equation}
Then one verifies that
\begin{itemize}\item[(a)] $R$ is analytic in $\mathbb C\setminus \Sigma_R$, with $\Sigma_R$ as given in Figure
\ref{figure: contour R algebr},
\item[(b)] for $\zeta\in\Sigma_R$, we have $R_+(\zeta)=R_-(\zeta)v_R(\zeta)$, where $v_R(\zeta)=I+\bigO(|t|^{-3/2})$ in the limit where $t\to -\infty$ or where $t\to +\infty$ and $s$ remains bounded away from $[-2\sqrt{3},\frac{2\sqrt{15}}{27}]$,
\item[(c)] as $\zeta\to\infty$, we have $R(\zeta)=I+\bigO(\zeta^{-1})$.
\end{itemize}
This is a small-norm RH problem which can be solved by a series expansion \cite{DKMVZ1}, and it follows from this standard procedure that \begin{equation}
R(\zeta;x,t)=I+\bigO(|t|^{-3/2}),
\end{equation}
uniformly in $\zeta$ as $t\to -\infty$ and also as $t\to +\infty$ with $s$ bounded away from the interval $[-2\sqrt{3},\frac{2\sqrt{15}}{27}]$.
Furthermore as $\zeta\to\infty$ we can expand $R$:
\begin{equation}\label{R infty}
R(\zeta)=I+\frac{R_1}{\zeta}+\bigO(\zeta^{-2}),\qquad\mbox{ as
$\zeta\to\infty$,}
\end{equation}
which implies that
\begin{equation}
R_1(x,t)=\bigO(|t|^{-3/2}).
\end{equation}
Thus by (\ref{def C1}) we have
\begin{equation}
B_1=\frac{z_0}{4}\sigma_3+\bigO(|t|^{-3/2}),
\end{equation} which results in the asymptotics (\ref{alg as y}) for $y$ by (\ref{yB}).

\section{Elliptic region}\label{section: ell}
In this section we consider the case where $t>0$ and $s\in
(-2\sqrt{3},\frac{2\sqrt{15}}{27})$. The construction done in the
previous section fails in this case because Proposition \ref{prop g
algebr} does not hold, which would result in exponentially growing
jump matrices for $T$ at certain parts of the contour. In order to
prevent this, we are forced to modify the $g$-function and afterwards to open lenses not only along $(-\infty,
z_0)$ but also on an additional interval.
\subsection{Construction of the $g$-function and modulation equations}
We search for a $g$-function in the form
\begin{equation}\label{g elliptic}
g(\zeta)=\frac{1}{30}\int_{\beta_1}^{\zeta}(z-\beta_3)^{1/2}(z-\alpha)(z-\beta_2)^{1/2}(z-\beta_1)^{1/2}dz,
\end{equation}
analytic in $\mathbb C\setminus(-\infty, \beta_1]$ and positive for
$\zeta>\beta_1$, where
\begin{equation}\label{def alpha}
\alpha=-\frac{1}{2}(\beta_3+\beta_2+\beta_1),
\end{equation}
with $\beta_3\leq \alpha\leq \beta_2\leq \beta_1$ real and depending on $s$ but not on $\zeta$.
As $\zeta\to\infty$, the $g$-function can be expanded as follows
\begin{equation}\label{expansion g elliptic}
g(\zeta;x,t)=\frac{1}{105}\zeta^{7/2}+c_1\zeta^{3/2}+c_2\zeta^{1/2}+d_1t^{-7/4}\zeta^{-1/2}+\bigO(\zeta^{-3/2}),
\end{equation}
with
\begin{align}
&\label{c1}c_1=-\frac{1}{360}((\beta_3+\beta_2+\beta_1)^2+2(\beta_3^2+\beta_2^2+\beta_1^2)),\\
&\label{c2}c_2=\frac{1}{360}((\beta_3+\beta_2+\beta_1)^3-4(\beta_3^3+\beta_2^3+\beta_1^3)).
\end{align}
Now we want to choose $\beta_3$, $\beta_2$, and $\beta_1$ in such a way that
\begin{equation}\label{asymptotics g2 elliptic}
    t^{7/4}g(\zeta;s)=\theta(t^{1/2}\zeta;x,t)+d_1\zeta^{-1/2}+\bigO(\zeta^{-3/2}),
    \qquad \mbox{as $\zeta\to\infty$.}
\end{equation}
This is true if
\begin{align}
&\label{mod1}(\beta_3+\beta_2+\beta_1)^2+2(\beta_3^2+\beta_2^2+\beta_1^2)=120,\\
&\label{mod2}(\beta_3+\beta_2+\beta_1)^3-4(\beta_3^3+\beta_2^3+\beta_1^3)=360s.
\end{align}
Furthermore, in order to be able to do a steepest descent analysis, we also require
\begin{equation}\label{mod3}
\int_{\beta_3}^{\beta_2}\sqrt{\xi-\beta_3}(\xi-\alpha)\sqrt{\beta_2-\xi}\sqrt{\beta_1-\xi}d\xi=0,
\end{equation}
or in other words $g_\pm(\beta_2)=g_\pm(\beta_3)$. It will become clear later on why we need to impose the latter condition.
We look for solutions to the modulation equations (\ref{mod1})-(\ref{mod3}) for which $\beta_3\leq \alpha\leq \beta_2\leq \beta_1$.
For two special values of $s$, the system of equations can be solved easily.
The first one corresponds to the confluent case where $\beta_3=\beta_2=\alpha$. Then (\ref{mod3}) is automatically satisfied, and (\ref{mod1})-(\ref{mod2}) are solved uniquely by \begin{equation}\label{modPII}
\beta_3=\beta_2=\alpha=-\sqrt{3}, \qquad \beta_1=4\sqrt{3}, \qquad s=-2\sqrt{3}.
\end{equation}
In the case where $\beta_2=\beta_1$, we have the unique solution
\begin{equation}\label{moddisc}
\beta_3=-\frac{4}{3}\sqrt{15}, \qquad \alpha=-\frac{1}{3}\sqrt{15}, \qquad \beta_2=\beta_1=\sqrt{15}, \qquad s=\frac{2\sqrt{15}}{27}.
\end{equation}
Those confluent cases correspond exactly to the values of $s$ which
are at the border between the algebraic and the elliptic region. For
$s$ in between the two critical values $-2\sqrt{3}$ and
$\frac{2\sqrt{15}}{27}$, the equations (\ref{mod1})-(\ref{mod3}) are
solvable \cite{Potemin}. Note that the different variable
$z=\frac{s}{\sqrt{6}}$ was used in \cite{Potemin}.

Before proceeding with the RH analysis, let us write down some useful properties of the $g$-function, relying on (\ref{g elliptic}) and (\ref{mod3}):
\begin{align}
&\label{prop g1}g_+(\zeta)+g_-(\zeta)=0,&\mbox{for $\zeta\in(-\infty, \beta_3)\cup(\beta_2, \beta_1)$},\\
&\label{prop
g2}g_+(\zeta)-g_-(\zeta)=2g_+(\beta_2)=2g_+(\beta_3)=-i\Omega,&\mbox{for
$\zeta\in(\beta_3, \beta_2)$},
\end{align}
with
\begin{equation}
\Omega=\frac{1}{15}\int_{\beta_2}^{\beta_1}\sqrt{\xi-\beta_3}(\xi-\alpha)\sqrt{\xi-\beta_2}\sqrt{\beta_1-\xi}d\xi.
\label{Omega}
\end{equation}

\subsection{Normalization of the RH problem and contour deformation}
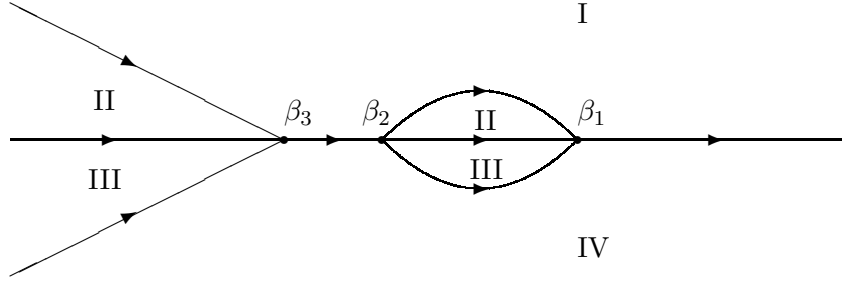
\begin{figure}[t]
\begin{center}
    \setlength{\unitlength}{1.3truemm}
    \begin{picture}(80,33)(10,25)
    \put(60,52){I}
    \put(10.5,43){II}
    \put(10,35){III}
    \put(49.5,41){II}
    \put(49,36){III}
    \put(60,28){IV}
        \put(60,40){\thicklines\circle*{.8}}
        \put(40,40){\thicklines\circle*{.8}}
        \put(30,40){\thicklines\circle*{.8}}
        \put(60,42){$\beta_1$}
        \put(38,42){$\beta_2$}
        \put(30,42){$\beta_3$}
        \put(30,40){\line(-2,1){28}}
        \put(30,40){\line(-2,-1){28}}
        \put(60,40){\line(-2,0){58}}
        \put(13,40){\thicklines\vector(1,0){.0001}}
        \put(51,40){\thicklines\vector(1,0){.0001}}
        \put(60,40){\line(1,0){27}}
        \put(75,40){\thicklines\vector(1,0){.0001}}
        \put(36,40){\thicklines\vector(1,0){.0001}}
        \put(15,47.4){\thicklines\vector(2,-1){.0001}}
        \put(15,32.6){\thicklines\vector(2,1){.0001}}
        \put(51,45){\thicklines\vector(1,0){.0001}}
        \put(51,35){\thicklines\vector(1,0){.0001}}
        \qbezier(40,40)(50,50)(60,40)
        \qbezier(40,40)(50,30)(60,40)
    \end{picture}
    \caption{The jump contour $\Sigma$.}
    \label{figure: contour elliptic}
\end{center}
\end{figure}
We will now use the $g$-function to normalize the RH problem at infinity in a suitable way. We define $T$ in such a way that it has jumps on a lens-shaped contour as shown in Figure \ref{figure: contour elliptic}:
\begin{equation}
T(\zeta; x,t):=\begin{pmatrix}1&0\\d_1t^{1/4}&1\end{pmatrix}\times\
\begin{cases}
Y_1(t^{1/2}\zeta; x,t)e^{t^{7/4}g(\zeta;x,t)\sigma_3},&\mbox{ in region I},\\
Y_2(t^{1/2}\zeta; x,t)e^{t^{7/4}g(\zeta;x,t)\sigma_3},&\mbox{ in region II},\\
Y_3(t^{1/2}\zeta; x,t)e^{t^{7/4}g(\zeta;x,t)\sigma_3},&\mbox{ in region III},\\
Y_4(t^{1/2}\zeta; x,t)e^{t^{7/4}g(\zeta;x,t)\sigma_3},&\mbox{ in region IV},
\end{cases}
\end{equation}
with $d_1$ given by (\ref{expansion g elliptic}).
Then $T$ satisfies the following RH problem.
\subsubsection{RH problem for $T$}
\begin{itemize}
\item[(a)] $T$ is analytic in $\mathbb{C}\setminus\Sigma$, where $\Sigma$ is as shown in Figure \ref{figure: contour elliptic}.
    \item[(b)] $T$ satisfies the following jump relations on
    $\Sigma$,
    \begin{align*}
        \label{RHP Phi2: b1}
        T_+(\zeta)&=T_-(\zeta)
            \begin{pmatrix}
                1 & e^{-2t^{7/4}g(\zeta;x,t)} \\
                0 & 1
            \end{pmatrix},&\mbox{for $\zeta>\beta_1$,}\\[1ex]
        T_+(\zeta)&=T_-(\zeta)
            \begin{pmatrix}
                1 & 0 \\
                e^{2t^{7/4}g(\zeta;x,t)} & 1
            \end{pmatrix},&\mbox{for $\zeta$ off the real axis,}\\[1ex]
        T_+(\zeta)&=T_-(\zeta)
            \begin{pmatrix}
                0 & 1 \\
                -1 & 0
            \end{pmatrix},&\hspace{-15cm}\mbox{for $\zeta\in (-\infty, \beta_3)\cup(\beta_2, \beta_1)$,}\\[1ex]
        T_+(\zeta)&=T_-(\zeta)
            \begin{pmatrix}
                e^{-it^{7/4}\Omega} & e^{-t^{7/4}(g_+(\zeta)+g_-(\zeta))} \\
                0 & e^{it^{7/4}\Omega}
            \end{pmatrix},&\mbox{for $\zeta\in (\beta_3, \beta_2)$.}
    \end{align*}
    \item[(c)]  $T$ has asymptotics of the following form as
    $\zeta\to\infty$,
    \begin{equation}\label{RHP Phi2: c}
        T(\zeta)=\left(I+B_1(x,t)\zeta^{-1}+\bigO(\zeta^{-2})\right)
        t^{-\frac{1}{8}\sigma_3}\zeta^{-\frac{1}{4}\sigma_3}N.
    \end{equation}
\end{itemize}
Away from the branch points $\beta_3, \beta_2, \beta_1$, it is
straightforward to verify that the off-diagonal entries in the jump
matrices for $T$ decay on
$\Sigma\setminus((-\infty,\beta_3)\cup(\beta_2, \beta_1))$ as
$t\to\infty$, if the lenses are chosen sufficiently close to the
real axis.

\subsection{Outside parametrix}
Ignoring the small jumps and the branch points, we obtain the following RH problem.

\subsubsection*{RH problem for $P^{(\infty)}$}
\begin{itemize}
\item[(a)] $P^{(\infty)}$ is analytic in $\mathbb{C}\setminus(-\infty, \beta_1]$.
    \item[(b)] On $(-\infty, \beta_1)$, we have
    \begin{align}
        \label{RHP Pinfty: b1}
        P^{(\infty)}_+(\zeta)&=P^{(\infty)}_-(\zeta)
            \begin{pmatrix}
                0 & 1 \\
                -1 & 0
            \end{pmatrix},&&\mbox{for $\zeta\in (-\infty, \beta_3)\cup(\beta_2, \beta_1)$,}\\[1ex]
        \label{RHP Pinfty: b2}
        P^{(\infty)}_+(\zeta)&=P^{(\infty)}_-(\zeta)e^{-it^{7/4}\Omega\sigma_3},&&\mbox{for $\zeta\in (\beta_3, \beta_2)$.}
    \end{align}
    \item[(c)]  $P^{(\infty)}$ behaves at $\infty$ as
    \begin{equation}\label{expansion Pinfty ell}
    P^{(\infty)}(\zeta)=\left(I+D_1\zeta^{-1}+\bigO(\zeta^{-2})\right)
        t^{-\frac{1}{8}\sigma_3}\zeta^{-\frac{1}{4}\sigma_3}N.
    \end{equation}
\end{itemize}

Similar RH problems have been solved many times in the literature in
terms of $\theta$-functions and meromorphic differentials, see e.g.\
\cite{DVZ, DKMVZ1, Bleher, KuiMo}. The only minor difference in our case
is that $\infty$ is a branch point. This RH problem can be solved
explicitly using the Jacobi $\theta$-function and elliptic
integrals.  We will construct the parametrix explicitly and refer to
\cite{Bleher} for the construction in more general settings.

\medskip

In addition to the RH conditions stated above, we need to construct
the outside parametrix in such a way that
\begin{equation}\label{condition outside branch}
P^{(\infty)}(\zeta)=\bigO(|\zeta-\beta_j|^{1/4}), \qquad \mbox{ as
$\zeta\to \beta_j$, $j=1, 2, 3$}.
\end{equation} If this is not the case, the construction of local
parametrices later on would fail. $P^{(\infty)}$ has the form
\begin{equation}\label{outside ell}
P^{(\infty)}(\zeta)=\begin{pmatrix}1&0\\ d
t^{1/4}&1+\frac{c_1}{\zeta-\beta_2}\end{pmatrix}\frac{1}{\sqrt
2}t^{-\frac{\sigma_3}{8}}
\gamma(\zeta)^{-\sigma_3}N\begin{pmatrix}h(\zeta)&0\\0&\hat
h(\zeta)\end{pmatrix}.
\end{equation}
Here $h, \hat h$ are scalar functions we will determine below, and
\begin{equation}\label{def gamma}
\gamma(\zeta)=\left(\frac{(\zeta-\beta_1)(\zeta-\beta_3)}{\zeta-\beta_2}\right)^{1/4}.
\end{equation}
If we construct $h, \hat h$ in such a way that they are analytic in
$\mathbb C\setminus (-\infty, \beta_1]$ and that they satisfy the
jump conditions
\begin{align}
&\label{jump h1}h_+(\zeta)=h_-(\zeta)e^{-it^{7/4}\Omega},&\mbox{for
$\zeta\in
(\beta_3, \beta_2)$},\\
&\label{jump hath1}\hat h_+(\zeta)=\hat
h_-(\zeta)e^{it^{7/4}\Omega},&\mbox{for $\zeta\in
(\beta_3, \beta_2)$},\\
&\label{jump h2}h_\pm(\zeta)=\hat h_\mp(\zeta),&\mbox{for $\zeta\in
(-\infty, \beta_3)\cup(\beta_2, \beta_1)$},
\end{align}
it follows from (\ref{outside ell}) and (\ref{def gamma}) that the
jump conditions (\ref{RHP Pinfty: b1})-(\ref{RHP Pinfty: b2}) are
satisfied. Moreover if $h, \hat h$ admit expansions of the form
\begin{align}
& \label{h
as}h(\zeta)=1+h_1\zeta^{-1/2}+h_2\zeta^{-1}+\bigO(\zeta^{-3/2}),
\qquad \mbox{ as $\zeta\to\infty$},\\
&\label{hath as}\hat h(\zeta)=1 - h_1\zeta^{-1/2}+
h_2\zeta^{-1}+\bigO(\zeta^{-3/2}), \qquad \mbox{ as
$\zeta\to\infty$},\end{align} and if we take $d=h_1$, $P^{(\infty)}$
has the expansion (\ref{expansion Pinfty ell}) with
\begin{equation}\label{DF}D_{1,11}=\frac{\beta_3-\beta_2+\beta_1}{4}+h_2,
\qquad D_{1,12}=-h_1 t^{-1/4}.\end{equation} We will now construct
$h, \hat h$ explicitly. Let us consider the third Jacobi
$\theta$-function
\begin{equation}
\theta(z;\tau)=\sum_{m=-\infty}^{+\infty}e^{2\pi imz+\pi i\tau m^2},
\end{equation}which is symmetric in $z$ and has the periodicity properties
\begin{equation}\label{jump theta}
\theta(z+1;\tau)=\theta(z;\tau), \qquad \theta(z+\tau;\tau)=e^{-\pi
i\tau-2\pi iz}\theta(z;\tau).
\end{equation}
Now we let $h, \hat h$ be of the form
\begin{equation}\label{def h}
h(\zeta)=\frac{\theta(0)}{\theta(c)}
\frac{\theta(u(\zeta)+c)}{\theta(u(\zeta))}, \qquad \hat
h(\zeta)=\frac{\theta(0)}{\theta(c)}
\frac{\theta(c-u(\zeta))}{\theta(u(\zeta))},
\end{equation}
where
\begin{align} &\label{def u}u(\zeta)=\frac{1}{2C}\int_{\infty}^\zeta \frac{
dz}{\sqrt{(z-\beta_3)(z-\beta_2)(z-\beta_1)}},\\
&C=\int_{\beta_3}^{\beta_2} \frac{
dz}{\sqrt{(z-\beta_3)(z-\beta_2)(z-\beta_1)}}=\frac{2K(\sigma)}{\sqrt{\beta_1-\beta_3}},\\
&\tau=i\int_{-\infty}^{\beta_3}\frac{
dz}{C\sqrt{(\beta_3-\zeta)(\beta_2-\zeta)(\beta_1-\zeta)}}=\frac{iK'(\sigma)}{K(\sigma)},\\
&\sigma=\sqrt{\frac{\beta_2-\beta_3}{\beta_1-\beta_3}},
\end{align}
and $K$ is the complete elliptic integral of the second kind. We
then have $u_+(\zeta)\equiv -u_-(\zeta)\mod \mathbb Z$ for $\zeta\in
(-\infty, \beta_3)\cup (\beta_2, \beta_1)$, and using (\ref{jump
theta}) we obtain (\ref{jump h2}). For $\zeta\in(\beta_3, \beta_2)$
we have $u_+(\zeta)-u_-(\zeta)=\tau$. Using the second property in
(\ref{jump theta}), we obtain (\ref{jump h1})-(\ref{jump hath1}) if
$c=\frac{t^{7/4}\Omega}{2\pi}$, with $\Omega$ as in (\ref{Omega}).
Now we can calculate $h_1, h_2$ in (\ref{h as})-(\ref{hath as}),
which leads to
\begin{align}
&\label{h1}h_1=\frac{1}{C}\left(\frac{\theta'}{\theta}(0)-\frac{\theta'}{\theta}(c)\right),\\
&\label{h2}h_2=\frac{1}{C^2}\left(\frac{\theta''}{2\theta}(c)-\frac{\theta''}{2\theta}(0)+
\frac{\theta'^2}{\theta^2}(0)-\frac{\theta'}{\theta}(0)\frac{\theta'}{\theta}(c)\right).
\end{align}

\medskip

From the standard theory of $\theta$-functions \cite{Fay} it follows
that $\theta(u(\zeta))$ has its only zero at $\zeta=\beta_2$, so
that $P^{(\infty)}$ has no singularities other than the branch
points. As $\zeta\to \beta_3, \beta_1$, $h$ and $\hat h$ is bounded
and this leads to (\ref{condition outside branch}). As
$\zeta\to\beta_2$, $h(\zeta)$ and $\hat h(\zeta)$ are of order
$\bigO(\zeta-\beta_2)^{-1/2}$. Here we need to exploit the freedom
to choose $c_1$ in (\ref{outside ell}) in such a way that
(\ref{condition outside branch}) holds. The value of $c_1$ can be
computed easily but is unimportant for us. This completes the
construction of the parametrix.

\subsection{Local parametrices near $\beta_3$, $\beta_2$, and $\beta_1$}
The local parametrices near the branch points can be constructed in
the same way as in the algebraic case using the Airy function, we
again refer to \cite{DKMVZ1, FIKN, CV1} for details. We do not need
the precise form of the parametrices here, it is sufficient to have
the existence of parametrices satisfying the conditions
\begin{itemize}
\item[(a)] $P$ is analytic in a fixed neighborhood $U_j$ of $\beta_j$,
\item[(b)] $P$ satisfies exactly the same jump conditions than $T$ inside $U_j$,
\item[(c)] For $\zeta\in\partial U_j$, we have $P(\zeta)=P^{(\infty)}(\zeta)(I+\bigO(t^{-3/2}))$ as $t\to\infty$.
\item[(d)] $T(\zeta)P^{-1}(\zeta)$ is analytic at $\beta_j$.
\end{itemize}
It should be noted that the construction of the local parametrix
relies on (\ref{prop g2}) and thus indirectly on (\ref{mod3}).
Condition (\ref{condition outside branch}) is also crucial.

\subsection{Final transformation}

Define
\begin{equation}
R(\zeta;x,t)=\begin{cases}T(\zeta;x,t)P^{(\infty)}(\zeta)^{-1},&\mbox{ for $\zeta\in\mathbb C\setminus U$,}\\
T(\zeta;x,t)P(\zeta)^{-1},&\mbox{ for $\zeta\in U_1\cup U_2\cup U_3$.}
\end{cases}
\end{equation}
Then after a similar argument as in the algebraic case, we have
\begin{equation}
R(\zeta;x,t)=I+\bigO(t^{-3/2}),
\end{equation}
uniformly in $\zeta$ as $t\to\infty$ with $s\in (-2\sqrt{3}+\delta,\frac{2\sqrt{15}}{27}-\delta)$, $\delta>0$.
As $\zeta\to\infty$ we have
\begin{equation}
R(\zeta)=I+\frac{B_1(x,t)-D_1(x,t)}{\zeta}+\bigO(\zeta^{-2}),\qquad\mbox{ as $\zeta\to\infty$,}
\end{equation}
which implies that
\begin{equation}
B_1(x,t)=D_1(x,t)+\bigO(t^{-3/2}),
\end{equation}
and by (\ref{yB}) and (\ref{DF}),
\begin{equation}
y(x,t)=2D_{1,11}t^{1/2}-D_{1,12}^2t+\bigO(t^{-1/2})=\frac{\beta_3-\beta_2+\beta_1}{2}t^{1/2}+2h_2t^{1/2}
-h_1^2t^{1/2}+\bigO(t^{-1/2}).
\end{equation}
Using (\ref{h1})-(\ref{h2}), we obtain
\begin{equation}
y(x,t)=\frac{\beta_3-\beta_2+\beta_1}{2}t^{1/2}-
t^{1/2}C^{-2}(\log\theta)''(0) +t^{1/2}C^{-2}(\log\theta)''
(c)+\bigO(t^{-1/2}).
\end{equation}
By standard manipulations for $\theta$-functions and elliptic integrals, this leads to (\ref{expansion elliptic intro}).

\section{Critical asymptotics for $y$ and possible generalizations}\label{section: crit}
In this section, we indicate how the critical expansions (\ref{expansion P2}) and (\ref{expansion sol}) can be obtained.
Afterwards we will make some remarks about asymptotics for the Br\'ezin-Marinari-Parisi solutions to higher members of the Painlev\'e I hierarchy and about asymptotics for certain unbounded KdV solutions.

\subsubsection*{Painlev\'e II asymptotics}
For $s$ near $-2\sqrt{3}$, we can proceed as in the algebraic case,
see Section \ref{section: alg}, with some modifications. The first
problem is that Proposition \ref{prop g algebr} does not hold for
$\zeta$ near $-\sqrt{3}$. This leads to jumps on the lines
$z_0+e^{\pi\pm\theta_0}$ which are not uniformly close to the
identity matrix. In order to overcome this, we need to close lenses
again at the point $-\sqrt{3}$, so that we have a contour as in
Figure \ref{figure: contour elliptic} but with $\beta_3=\beta_2$.
Then the jumps will converge to the identity matrix as $t\to\infty$
except in a small neighborhood of $-\sqrt{3}<z_0$. Near this point,
we need to construct a local parametrix built out of
$\Psi$-functions associated with the Hastings-McLeod solution to
Painlev\'e II \cite{FN, IN}. This construction is essentially the
same as the one in \cite{CG2}. The calculations that finally lead to
(\ref{expansion P2}) are rather tedious, and as they are similar as
in \cite{CG2}, we do not think it is appropriate to include the
details in this paper.

\subsubsection*{Solitonic asymptotics}
For $s$ near $\frac{2\sqrt{15}}{27}$, we again proceed as in Section
\ref{section: alg}, but now Proposition \ref{prop g algebr} breaks
down near $\sqrt{15}>z_0$. There is no need to modify the jump
contour for $T$ here. It is however necessary to build a local
parametrix near $\sqrt{15}$. This time the parametrix has to be
constructed using Hermite polynomials, similarly as in \cite{CG3}.
The degree of the Hermite polynomials will depend on the value of
$\xi$ in (\ref{expansion sol}). If $\xi$ is close to a half positive
integer, a transition to Hermite polynomials of higher degree takes
place, and this requires a modified local parametrix. A long
calculation for which we refer to \cite{CG3} leads to
(\ref{expansion sol}).

\subsubsection*{Higher members of the Painlev\'e I hierarchy}
The Painlev\'e I hierarchy contains an infinite number of equations
${\rm P}_{\rm I}^{m}$ of order $2m$ with $m=1, 2 \ldots$ The
equation for $m=1$ is the Painlev\'e I equation $y_{xx}=x+6y^2$, and
the equation for $m=2$ is, up to a transformation $x\mapsto ax,
t\mapsto bt$ given by (\ref{PI2}). Br\'ezin, Marinari, and Parisi
\cite{BMP} considered not only the case $m=2$, but the general case
$m=2k$. They believed that for any $k\in\mathbb N$, there is a real
pole-free solution $y_k$ to the $2k$-th member of the hierarchy
which has asymptotics of the form
\begin{equation}\label{asymptotics y hierarchy}
    y_k(x)\sim \mp|c_kx|^{\frac{1}{2k+1}},
        \qquad\mbox{as $x\to\pm\infty$.}
\end{equation}
This conjecture was supported by Moore \cite{Moore} when he
considered the RH problem for the $2k$-th member of the hierarchy.
Although the general RH problem for the ${\rm P}_{\rm I}^{2k}$
equation has $4k+2$ Stokes multipliers, only three among them are
non-zero for the special solution under consideration: the one
corresponding to the positive real line and the ones corresponding
to the two anti-Stokes lines closest to the negative real line, all
three of them being equal to $1$ with the orientation as in Figure
\ref{figure: RHP Psi}. Adding $2k-1$ monodromy preserving time
parameters to the RH problem as in (\ref{definition: N theta}),
where $\theta$ now has a leading order term
$c\zeta^{\frac{4k+3}{2}}$, one can study long time/space double
scaling asymptotics as we did for $k=1$. Following the general
procedures of the Riemann-Hilbert analysis for Painlev\'e equations
\cite{FIKN}, one expects again regions of algebraic and elliptic
asymptotic behavior, but in addition also regions of hyperelliptic
behavior. The transitions between those regions would be interesting
to study as well, and might lead to more general Painlev\'e II
hierarchy asymptotics, a more general form of solitonic asymptotics,
and possibly also to critical asymptotics in terms of the special
solutions to the ${\rm P}_{\rm I}^{2j}$ equation with $j<k$.

\subsubsection*{Unbounded solutions to the KdV equation}
We already mentioned that the RH problem for ${\rm P}_{\rm I}^{2}$
generates solutions to the KdV equation by (\ref{def y}), also in
the case where the Stokes multipliers $s_0, \ldots, s_6$ depend on
$\zeta$. For generic choices of $s_j(\zeta)$, $y$ will have
poles at certain values of $x$ and $t$. However, for initial data
satisfying $y(x,0)=\bigO(|x|^{1/3})$ as $x\to \pm\infty$, the Cauchy
problem for KdV is well-posed \cite{Menikoff}. It would be
interesting to see if such solutions can be generated by choosing
appropriate Stokes multipliers $s_j(\zeta)$, and if those
solutions have asymptotic expansions similar to the ones for
$y(x,t)$.

\section*{Acknowledgements}
The author is grateful to T. Grava and B. Dubrovin for useful discussions and comments.
He is a Postdoctoral Fellow of the Fund for Scientific Research - Flanders (Belgium), and was also supported by
Belgian Interuniversity Attraction Pole P06/02, by the ESF program MISGAM, and by ERC Advanced Grant FroMPDEs.

\obeylines \texttt{Tom Claeys
    Cit\'e Scientifique - Laboratoire Painlev\'e M2
    F-59655 Villeneuve d'Ascq, FRANCE
E-mail: tom.claeys@math.univ-lille1.fr
}

\end{document}